# Towards International Relations Data Science:

# Mining the CIA World Factbook

---

University of Macedonia

Department of International and European Studies

Thessaloniki, Greece

Author

Panagiotis Podiotis, ies16079@uom.edu.gr

Register Number: ies/16079

Supervisor

Dr. Nikolaos Koutsoupias, nk@uom.edu.gr

Evaluation Committee

Dr. Dimitrios Vagianos, vagianos@uom.gr

## Abstract

This paper presents a three-component work. The first component sets the overall theoretical context which lies in the argument that the increasing complexity of the world has made it more difficult for International Relations (IR) to succeed both in theory and practice. The era of information and the events of the $21^{st}$ century have moved IR theory and practice away from real policy making (Walt, 2016) and have made it entrenched in opinions and political theories difficult to prove. At the same time, the rise of the "Fourth Paradigm – Data Intensive Scientific Discovery" (Hey et al., 2009) and the strengthening of data science offer an alternative: "Computational International Relations" (Unver, 2018). The use of traditional and contemporary data-centered tools can help to update the field of IR by making it more relevant to reality (Koutsoupias & Mikelis, 2020). The "wedding" between Data Science and IR is no panacea though. Changes are required both in perceptions and practices. Above all, for Data Science to enter IR, the relevant data must exist. This is where the second component comes into play. I mine the CIA World Factbook which provides cross-domain data covering all countries of the world. Then, I execute various data preprocessing tasks peaking in simple machine learning which imputes missing values providing with a more complete dataset. Lastly, the third component presents various projects making use of the produced dataset in order to illustrate the relevance of Data Science to IR through practical examples. Then, ideas regarding the future development of this project are discussed in order to optimize it and ensure continuity.

Overall, I hope to contribute to the "fourth paradigm" discussion in IR by providing practical examples while providing at the same time the fuel for future research.

## Contact Info, Code and Data Dissemination

Author: Panagiotis Podiotis | Email: ies16079@uom.edu.gr

LinkedIn: https://www.linkedin.com/in/panagiotis-podiotis/

GitHub Repository: https://github.com/Podipan/cia_world_factbook

Research Gate: https://www.researchgate.net/profile/Panagiotis_Podiotis

*-Dedicated to grandpa Antonis who taught me hard work, pride and laughter.*

# Table of Contents

# Chapter 1 – Introduction

## 1.1 Relevance to the field

*"In discipline after discipline . . . academics have all but lost sight of what they claim is their object of study"* (Ian, 2005, p. 2)*;* and the discipline of International Relations is no exception. Praising the advantages and the necessity of the field of International Relations to politics and society would be redundant in terms of time and paper when the field can only be strengthened when its weaknesses are realized and solutions are sought. In that spirit, the field of International Relations, either as a subfield of Political Science (Kouskouvelis, 2007, pp. 22-23) or even as Political Science itself (Carr, 1939), suffers from inherent weaknesses. These weaknesses are evident on both sides of the same coin. Namely, International Relations' theory[1] and application[2]. While the debate around IR's theoretical weaknesses is still going on, there is much wider consensus on its practical failures (Baron, 2014). Factors like the lack of experimentation data, the long term aspect of political outcomes, the difficulties in substantiating an opinion into a theorem and the varying schools of thought (Martin, 1972, p. 846) all make IR's role significantly harder in interpreting the world and forecasting the future. Make no mistake, this paper will not discuss the Great Debates (David, 2013) nor will it try to discredit IR theory. It merely serves to promote a practical multidisciplinary approach to International Relations.

Despite the fact that Complexity Sciences (Caws, 1963) and fields like System Dynamics Models, Data Mining and Quantitative methods (Unver, 2018) have been around for many decades, their application in IR has been relatively limited when compared to other sectors. Technical literature remains underdeveloped and most guides for the young IR student come from the field of Computer Science and non-academic[3] sources. Philip A. Schrodt in his paper "Forecasting Political Conflict in Asia using Latent Dirichlet Allocation Models" (Schrodt, 2011) presents a plethora of computer-powered IR-related projects which, as mentioned before, exist on high institutional levels. Such technical efforts to assist IR in describing, explaining or forecasting reality should be further democratized for IR students. While data science courses start to appear and increase in political sciences study programs, few political science students understand the potential or utilize the relevant tools. The ideas of the Fourth Paradigm as described in (Hey et al., 2009) are now more relevant than ever. While most aspects of human life become increasingly reliant and invested in the mining and use of data, international relations are falling behind. It is high time that the “fourth paradigm – data intensive scientific discovery” is re-examined in International Relations. It is time we give “Computational IR” (Unver, 2018) a chance.

[1] A non-exhaustive list of examples being the work of Helen Milner in "Review: International Theories of Cooperation among Nations: Strengths and Weaknesses" (Helen, 1992, p. 481) and the article of W. Julian Korab-Karpowicz "Political Realism in International Relations" (2010).
[2] Certain aspects of realism are discussed in Barak Obama's interview (Goldberg, 2016) for the Atlantic. Even though IR theory is not discussed in an official academic setting, the clutter surrounding political theory and actual practice becomes quite evident especially when additional other non-academic sources are reviewed.
[3] Mostly online forums, websites, video guides etc.

Events such as: The annexation of Crimea by Russia, BREXIT, the revolutions in the middle east[4] (Gadi et al., 2013), the election of Donald Trump[5] and the SARS-CoV-2 outbreak. Events like these highlight the modern challenges for the policy-maker and decision-taker. In the era of Information (Kremer & Müller, 2014), political intelligence (Fishman & Greenwald, 2015), within the big data and OSINT realms, assumes central role for leaders. Kenneth Waltz’s single picture prescriptions as painted in *Man, the State and War: A Theoretical Analysis* (Waltz, 1959, p. 300) seem to become synonymous with IR theory itself. A theory becoming increasingly unable to capture the chaotic modern world. This proposition can be illustrated through the continuous reexamination of realism (Korab-Karpowicz, 2010), one of the most recognized branches of IR theory. The *“…search for the inclusive nexus of causes”* (Waltz, 1959, p. 300) can become easier to materialize with the democratization of computational machines and algorithms. Solid theory and political practices can be boosted by capitalizing on data mining and analytics tools. It is this very ability to extract quantitative and qualitative data from societies which can be used for more effective policies (Koutsoupias & Mikelis, 2019), substantiated opinions and theory that makes the multidisciplinary cross-over between IR (Rosenberg, 2016) and Data Science so attractive and useful. Under no circumstances does this introduction aims to reduce the importance of IR and of related theories. It serves as a reminder that multidisciplinary knowledge and projects can lead to better policies and theory evaluation. I am aware to the fact that this new approach may ultimately fail, but considering the probable benefits it is definitely worth trying.

---

[4] Called by many “Arab Spring”. The appearance of the Islamic State brought the internet in the global spotlight as an invaluable asset to terrorism (Weimann, 2010), drawing a direct connection between the internet and national security of states.

[5] Not mentioned in a negative context but rather in terms of unpredictability and influence of the internet.

## 1.2 Structure, Aims and Objectives

The paper at hand consists of three components, one theoretical and two technical. The theoretical one, calls for the deeper use of Data Science in IR and for the strengthening of Computational International Relations. The other two technical components support the arguments of the theoretical one by creating, using and presenting actual tools and methods.

This paper can thus be illustrated as:

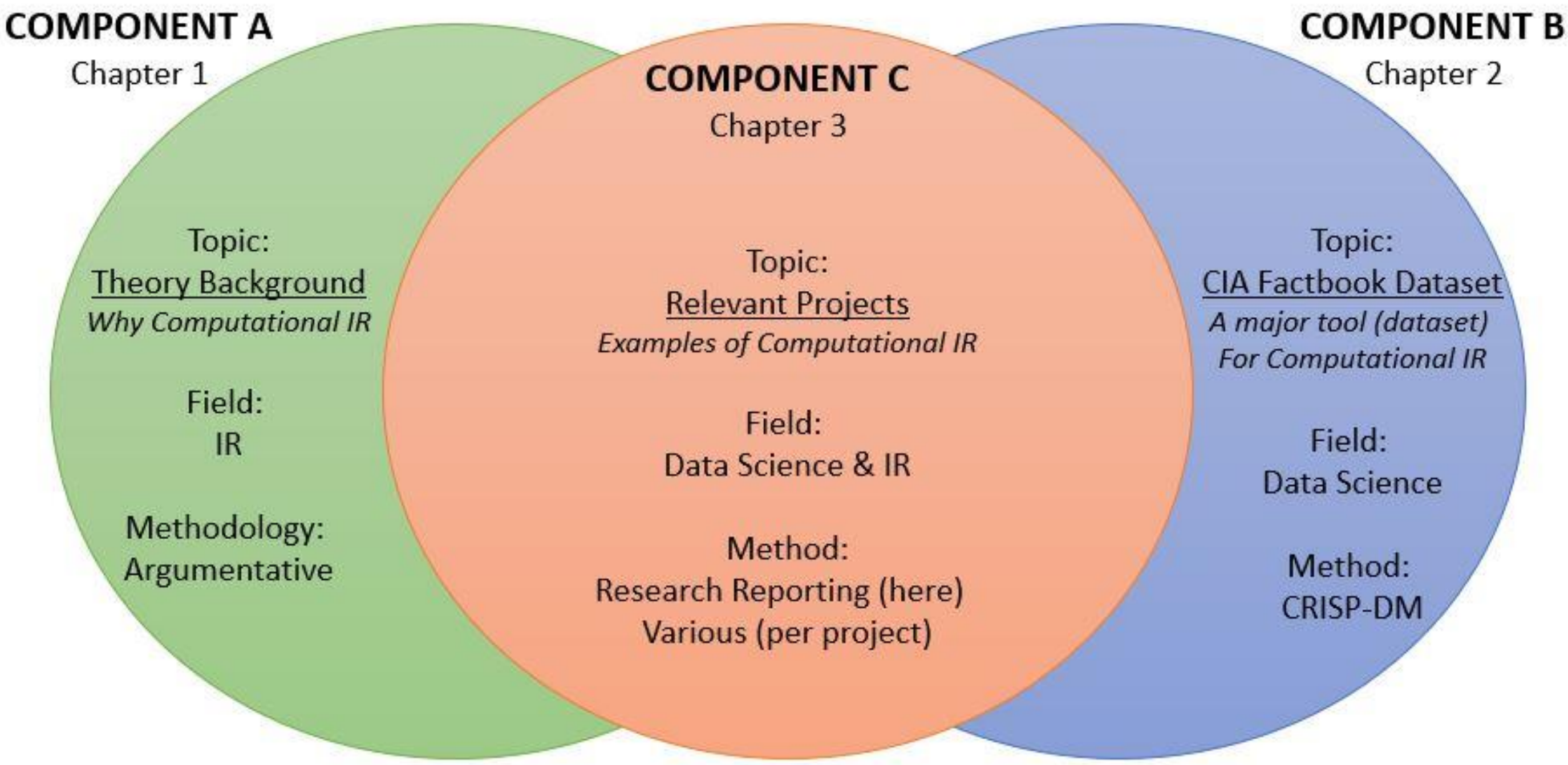

*Figure 1 - Paper's Structure*

Overall, the author aims to 1) provide new literature for the field of International Relations Data Science by raising the awareness; 2) create a high-quality dataset which will serve as groundwork for future research; and 3) develop his knowledge and abilities on Data Science by exploring various algorithms and concepts. In order for the above aims to be met, it is crucial that: a) This application is effectively placed within the International Relations context; b) Methodology and executed steps are presented in a concise and detailed manner; c) Additional uses of the data and tools produced are discussed in order to ensure continuity.

## 1.3 Methodology

Considering the ideas discussed in section 1.1 Relevance to the field, the present work can be regarded as multidisciplinary in nature. Specifically, an explanatory data-science project standing within the theoretical IR cosmos.

The reader should expect a variety of methodologies within this paper. This is due to the fact that this paper has both technical and theoretical aspects. The three-component structure presented in 1.2 Structure, Aims and Objectives plays a crucial in role in that regard.

The theoretical part is argumentative and heavily influenced by the ideas of the "Fourth Paradigm" (Hey et al., 2009), "Computational IR" (Unver, 2018) and by the experiences of the late 2010s. Its aims are mostly met in 1.1 Relevance to the field.

Moving on to the two technical components, the Chapter 3 – Relevant Projects part presents some findings of other projects which were extracted from the CIA World Factbook dataset by using Statistical, Natural Language Processing and Machine Learning tools. For more details on the methodologies and methods of these projects, the reader is strongly advised to refer to the relevant papers directly.

The CIA World Factbook Dataset creation, which is the center of gravity of this paper, was carried out using a simplified version of CRISP-DM[6]. The selection criteria and other details are presented below. Due to the lack of data-centered methodologies in IR, Data-Science methodologies[7] were examined. CRISP-DM was selected and modified in order to better serve the requirements of this project (for a comprehensive summary of the project's pipeline see Appendix A. Program Pipeline). Modifications conducted mostly refer to removing certain business and model deployment steps. Such steps are irrelevant to this project since it is placed within a large corporation or institution with bureaucracy nor is it ultimately producing a model for use outside of its narrow scope.

[6] Summarized effectively in (Shearer, 2000) and (Hipp & Wirth, 2000).

[7] A variety of online and other sources were examined with the initial references being: (Azevedo & Santos, 2008), (Martínez-Plumed, et al., 2019) and (Mariscal, Marbán, & Fernández, 2010).

The selection criteria which led to the CRISP-DM methodology were:

1. Fit for the present project. This project is mostly about data-mining and processing rather than advanced modeling.
2. Widely used, proven and accepted (Gregory, 2014).
3. Highly adaptable.
4. Extensive bibliography.
5. Easily interpretable.
6. Multidisciplinary character.

Under these criteria, the aforementioned modified version of CRISP-DM was selected as the ideal[8] methodology.

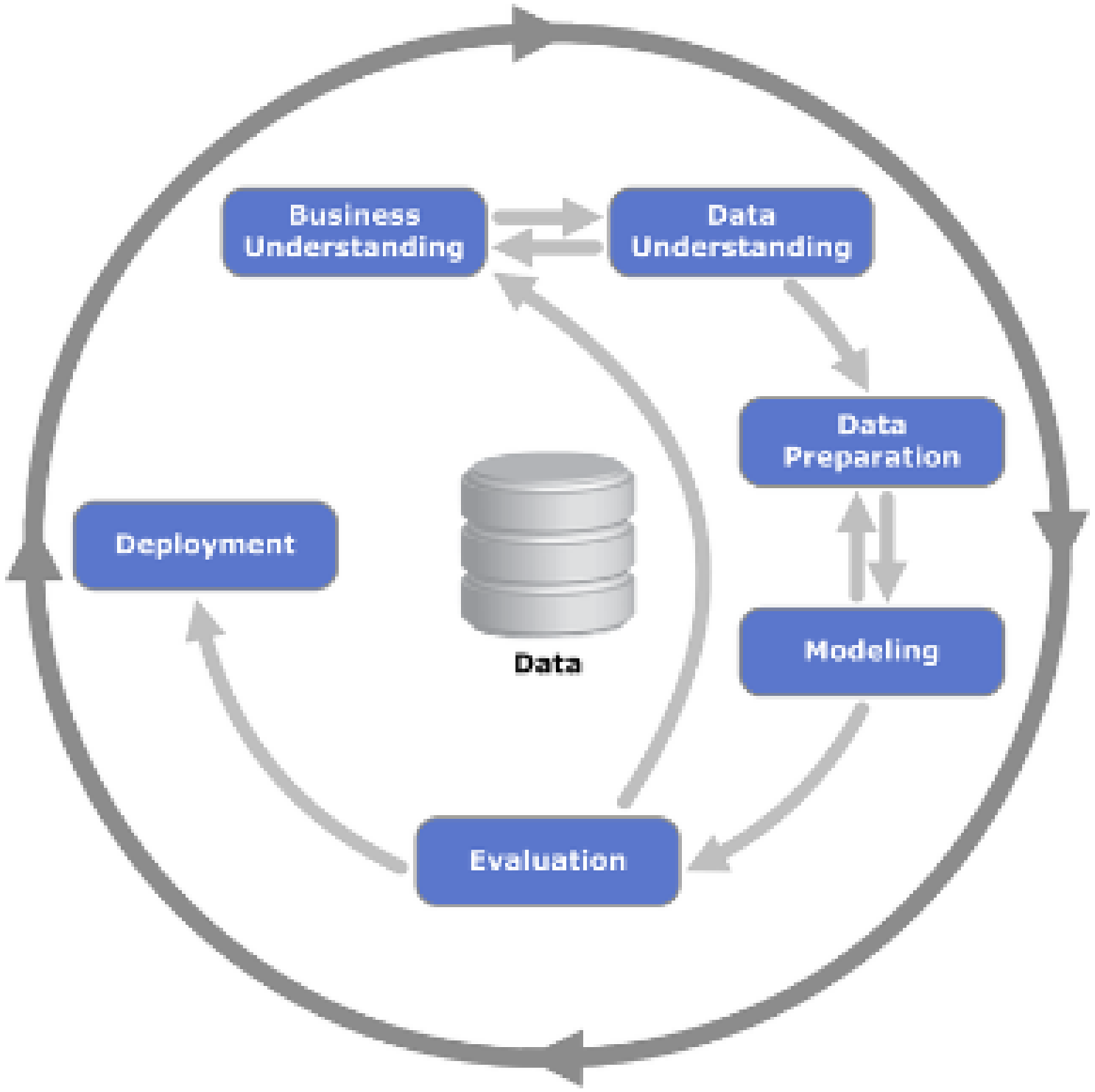

*Figure 2 - The CRISP-DM reference model. From  (Wikipedia contributors, 2019), cross validated with (Shearer, 2000)*

[8] The author is aware that CRISP-DM may be ideal but not optimal. There is a great number of robust methodologies but an evaluation of each one not only falls outside the scope of this paper but is also challenging in terms of time.

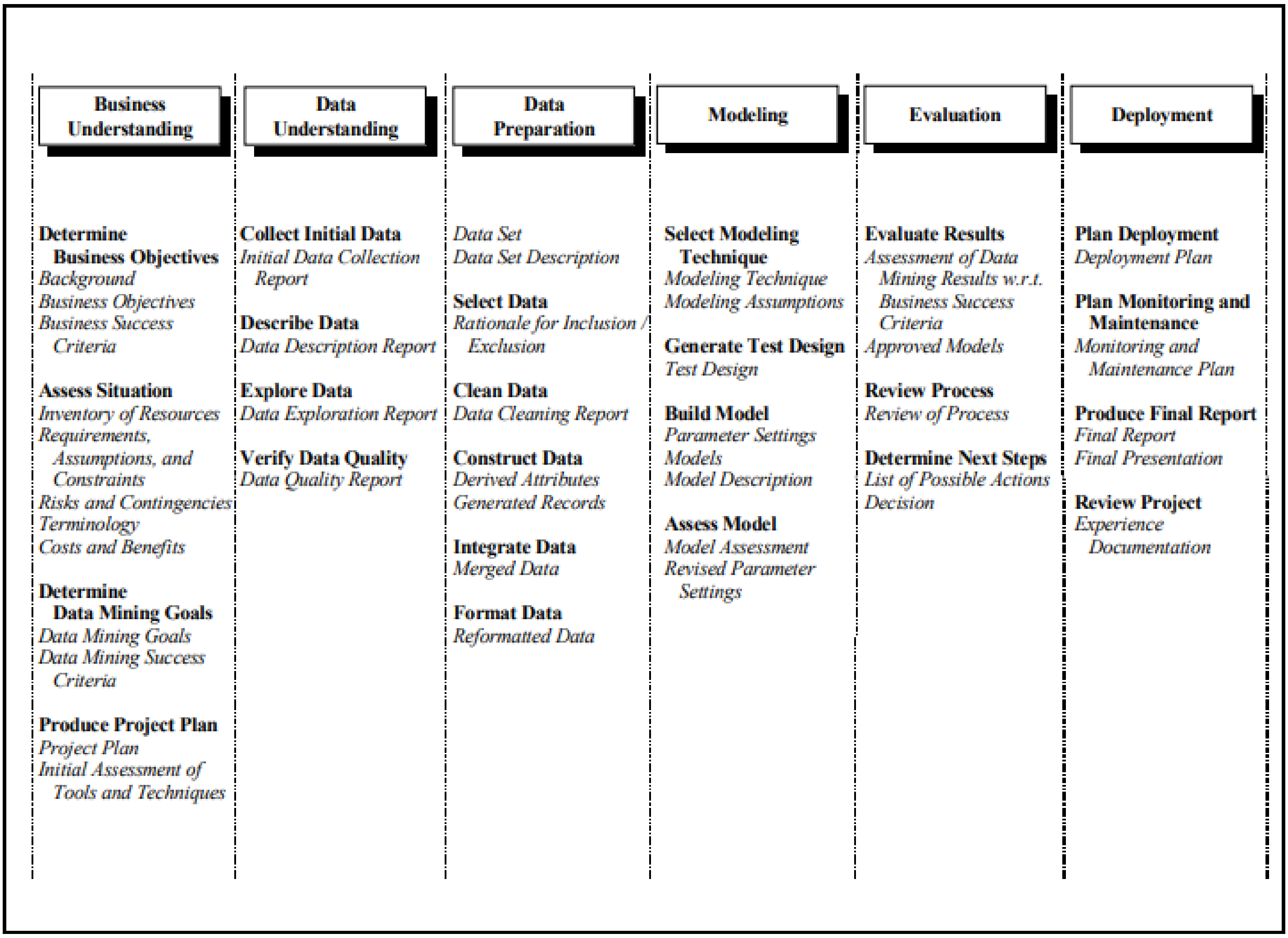

*Figure 3 - CRISP-DM tasks and outputs. (Hipp & Wirth, 2000)*

A more technical presentation of the various python scripts is laid out in the project plan provided below. The summarized version can be found in Appendix A. Program Pipeline.

| Project Plan | | | | |
|---|---|---|---|---|
| Phase | Description | Activities (Descending) | Output | Tools |
| 1 | Primary Data Mining - CIA World Factbook Extraction | Data Download<br>Data Inspection<br>Dataframe Creation<br>Data type processing<br>Dataframe Filling | Dataframe v1 (Raw Factbook) | Those described in 2.1.2 Inventory of Resources along with libraries pandas, numpy, matplotlib |
| 2 | Size Reduction | Non-state entries removal<br>Column concatenation (terrorism and borders)<br>Missing not at Random (MNAR) filling<br>Column removal (if empty >95%) | Dataframe v2 (Preprocessed) | |
| 3 | Data transformation | Textual data to numerical or labels. | Dataframe v3 (Transformed) | |
| 4 | Feature Engineering | Encoding of label data to binary (dummy variables) with OneHot Encoding. | Dataframe v4 (Extended) | |
| 5 | Model Selection and tuning. | Literature exploration<br>Model Testing<br>Evaluation | List of Machine Learning models, parameter, metrics etc. | Existing academic, scientific and online literature. |
| 6 | Model Implementation | Simple OLS Linear Regression<br>Multiple Ridge Regression<br>Random Forest Regression | Dataframe v5 (Imputed-Finalized) | Same as phases 1-4 plus sklearn for ML. |

*Table 1 - Project Plan*

Specific technical details are discussed in each respective section. Dataframes (for a summary of all project's Dataframes see Appendix B. Overall Dataframe Versions, Processing, Shape) generated by steps 1-3 were evaluated comparatively to the online Factbook in order to verify that minimal data loss and distortion had occurred. The Dataframe generated from Step 4 was compared (human) to Dataframe v1 in order to observe whether semantics had been altered during the transformation. Step 5 was iterated multiple times and outcomes were compared. After the optimal models and parameters were found, step 6 was executed effectively forming the last version of the Dataframe (v5). After the publication of this paper, the code will be uploaded on the web. Comments and suggestions of the online community will be taken under consideration and the code will hopefully keep evolving on the public domain.

## 1.4 Data Source (The CIA World Factbook)

With respect to the IR background of this paper, it was obligatory that data describing states, the largest actors of the International System, were found. Moreover, the author of the original data must enjoy a high degree of validity and be a recognized authority. In that sense, considering the US as one of the most important global policy powers and the CIA as one of the most recognized intelligence agencies, the CIA World Factbook seemed ideal. Furthermore, the policy[9] governing the use of the Factbook data gives flexibility to the researcher. Last but not least, the inquisitive nature of this paper along with the author's familiarity[10] with the Factbook also played an important role in its selection.

The CIA World Factbook is a US Central Intelligence Agency publication, firstly published in 1981, which: *"provides information on the history, people and society, government, economy, energy, geography, communications, transportation, military, and transnational issues for 267 world entities"* (CIA, 2019, p. About). The Factbook is integrated into CIA's website www.cia.gov and is updated regularly. Printed versions are also available for download. All of the data are open-source and can be copied freely as stated in (CIA, 2019, p. Copyrights and Contributors). Factbook's content is provided to the CIA by a number of US governmental institutions.

---

[9] I would like to acknowledge that the copyrights policy of the CIA World Factbook should constitute an example for all state organizations handling data. This paper would have not been possible if a different policy was active.
[10] The author has been using the CIA World Factbook for a plethora of academic papers and research.

# Chapter 2 – The CIA World Factbook Dataset

## 2.1 Business Understanding

### 2.1.1 Background

Background information regarding the importance and need for this project are provided in 1.1 Relevance to the field. Existing IR datasets[11] are limited in scope usually focused on a specific sector in great depth. Multidomain datasets in regard to IR are scarce and data are usually of low validity. In an effort to alter the situation described above, an objective to create a complete, reliable and easy to use dataset for the International Relations study and policy has been set. An optimization approach was adopted and will continue to unravel even after this paper.

### 2.1.2 Inventory of Resources

The project's inventory of resources has various disparities. While the human capital and hardware capabilities are limited, the amount of data and software tools are rather robust and extensive. In our case, the CIA World Factbook has been selected as the source of data. For the rationale behind the choice of this source, please see 1.4 Data Source (The CIA World Factbook). The Factbook offers a great amount of data categories covering more than all countries by providing metrics, text and historic data across 12 sectors (History, Geography, People and Society, Government, Economy, Energy, Communications, Military and Security, Transportation, Terrorism, Transnational Issues). This great wealth of information can be harvested and processed with the widely used programming language python which comes in hand with various industrial-grade software libraries. Specifically, the Project was executed with the use of Python 3.7 & 3.8 (Python Foundation, 2001) programming language mostly within the JetBrains PyCharm Community Edition 2019.2.2 x64 (JetBrains, 2000) Integrated Development Environment. Few visualization and exploration tasks were carried out in Kaggle (Kaggle, n.d.). Various libraries were used, more details in the respective parts of the paper. All data and libraries were contained in a project-specific virtual environment. Generated data were saved in pickle format in order for data types to be maintained and excel (.xlsx) format for human exploration. All software tools used are available for free use or licensed to the device used.

[11] Such as World Bank's, national census/statistics agencies or datasets on conflict.

All tools performed as designed during every iteration of activities. All libraries used were selected based on the following criteria:

1. Presence in relevant literature (fame).
2. Robustness.
3. Extensive documentation.
4. Ease of use considering the non-technical background of the author.

### 2.1.3 Assumptions, Limitations, Constraints

The assumptions regarding this project are:

1. CIA World Factbook data are expected to be of high confidence and low bias.
2. Some data were determined to be Missing not at Random at the discretion of the author (see Appendix D. Columns Assumed MNAR)
3. Some assumptions are made while converting textual columns to numerical or labels (see Appendix E. Columns Generated from Textual - Assumptions)
4. The data produced or transformed by each algorithm were manually compared to the data of the previous phase in order to determine quality of implementation. Thorough manual comparisons, sometimes pc assisted, were assumed to provide with a good picture of overall algorithm effectiveness.

The constrains and limitations of the project are:

1. Existing academic theory is limited in regards to Data Science applications for IR.
2. There is a lack of academic technical guides for IR students.
3. IR students tend to have weak technical literacy.
4. There is a general lack of consensus and axioms regarding machine learning model evaluation metrics. Most of the related knowledge is empirical.
5. There are no methodologies covering data science in International Relations.
6. The gap between IR theory, actual policy making and data science remains vast.

7. Information such as units of measure is provided separately in the CIA website thus an important amount of manual labour is required.
8. Lack of standardization in textual data makes conversion to labels difficult. Vectorization could be an effective alternative, to be tested in future versions of the code.

### 2.1.4 Goals and success criteria

This project aims to extract the maximum amount of Data from the CIA World Factbook to a panda's Dataframe and then calculate missing values with machine learning. The dataframe must contain refined data, be easily interpretable and ready to be used by individuals irrelevant to this paper. Missing values of the dataset must be imputed accurately in order for a more complete-dataset to be produced. It is thus a project consisting of three major parts. 1) Data Extraction 2) Data Processing 3) Missing values imputation (Prediction).

The creation of an easy to use and accurate Dataframe representing the CIA World Factbook with high confidence is the primary success criterion. The referencing and usage of the dataset in the public domain will constitute the ultimate success evaluation metric along with the relative feedback. Adaptability of the code to future versions of the Factbook is also highly desired.

## 2.2 Data Understanding

### 2.2.1 Initial Data Collection

Data were downloaded though the official CIA World Factbook website[12] thus no web scrapping was needed. The downloaded file (.zip) not only contained the data for all Factbook entries but also included pictures, appendices, fonts etc. In our case, only the files containing entry data were used. Such data were found in .html and .json files. Moreover, .json files contained per-entry data while the .html files contained per-field data covering all Factbook entries. In essence, both .json and .html files contained the scrapped online Factbook pages in HTML, same data but different perspectives. All data, including numbers, are by default handled as strings. Numbers needed to be defined as such.

[12] https://www.cia.gov/library/publications/download (accessed 15/10/19)

### 2.2.2 Data Extraction and Preprocessing

Moving on, the structure of the downloaded files was explored comparatively to the online version. It was discovered that the Factbook is structured in the following order (Descending):

**Examples in parentheses.*

Entity[13] (Mexico) --> Category (Economy) --> Field (GDP) --> Subfield[14] (GDP per capita)

A single entity can contain up to 12 Categories. The categories of a single entity can contain a maximum of 186 fields. Each field contains a minimum of one subfield. Subfields have varying structures and can be of one out of four types: Numerical, Textual, Historical or Grouped. In order for the maximum amount of data to be captured and for data types to be preserved and defined, it was essential that subfield data were captured. The need to capture all data at the lowest level possible along with the lack of structure resulted in the failure of xpath approaches which failed to adapt sufficiently to the structure of the Factbook. Moreover, html reading functions and existing algorithms did not perform as intended due to the fact that the data downloaded were in distorted HTML format, a byproduct of the scrapping conducted by the CIA at the first hand. With that in mind, a custom-made approach was developed.

At first, an empty panda's DataFrame was created in order to accommodate all extracted data. Then, all 268 json files representing individual Factbook entities were opened iteratively using python OS functions, all characters were converted to lowercase and then the subfields of each entity were parsed using regular expressions and string manipulation methods and functions. The data extraction began from the category level. Specifically, "<div class=\\\"category " was used as splitting point for categories. Next, each category was split into Fields using "<div class='category_data subfield" as reference point. Each field was searched for contained subfields using ""subfield-". The title of each field/subfield was found and the relative column was created on the dataframe. Finally, the field's/subfield's data were processed, defined and appended in accordance to the following steps:

---

[13] Includes US recognized: States, territories, geographic regions, etc. According to (CIA, 2019, p. Appendix D) all geopolitical entities forming the CIA World Factbook fall under FIPS PUB 10-4 and with respect to the recommendations of the Board on Geographic Names (BGN). *"Territorial occupations/annexations not recognized by the United States Government are not shown on US Government maps."* (CIA, 2019, p. FAQs)

[14] Contrary to other structural elements, Subfields are not present in every Field.

Numerical fields/subfields were first stripped of any text appearing as “note”. Data were then filtered through a custom-made function[15]. All numeric columns on the dataframe were labeled by the word “num” at the beginning of their title.

Textual fields/subfields were first cleaned of any remaining html and country rankings which are redundant information due to the fact that it is very easy to reproduce them in our dataframe simply by sorting the related numeric column. Textual columns were labeled in the dataframe by the word “txt” at the beginning of their title.

Historical fields/subfields were treated as numeric with the only exception being that only the latest[16] available data were preserved. This was achieved by splitting and then appending data in lists only to keep the most recent. Historical columns were labeled in the dataframe by the word “num” at the beginning of their title and “hist” at the end.

Grouped fields/subfields were split into groups using "subfield-name\\\">" as a reference point and were then treated as numeric. Grouped columns were labeled in the dataframe by the word “num” at the beginning of their title due to the numerical type of their data.

After all files had been processed, empty columns were dropped from the DataFrame and ("N/A", "na", "na%", "nan", "n/a", "$na")[17] values were replaced with numpy’s nan (np.nan). The dataframe was saved in pickle (.pkl) format for future processing and as an excel (.xlsx) file for visual inspection and ease of accessibility. (For an overview of project’s Dataframe versions see Appendix B. Overall Dataframe Versions, Processing, Shape). The resulting Dataframe (1st version) has 268 rows, each one representing a single Factbook entity and 536 columns referring to unique subfields[18]. For the indexing of rows, country codes were used and for columns naming composite titles.

[15] Function searched for million/billion/trillion words appearing next to the number, applied the respective multiplication and then removed all text returning back a float value. Inputs and outputs were visually inspected for errors and corrections to the code were implemented until none could be found.

[16] At first it may seem that this approach reduces the timeseries forecasting capabilities of the dataset. Such concerns will be addressed in the conclusions chapter.

[17] These strings are found throughout the dataset where no data are provided for the respective subfield/field.

[18] In case a field contains only one subfield then the name of the field is adopted by the subfield and as a consequence the Dataframe’s column.

Every title consists of three parts. Parts are separated by a single whitespace. Part 1 indicates the contained data type, part 2 indicates the category and field and part 3 indicates the subfield. The only exceptions to above nomenclature rule are columns: “Country Code”, “txt Country Name” and “lbl Region”. These columns draw their content directly from the metadata and not from the main body of the HTML code. (For a summary of column nomenclature see Appendix C. Column Nomenclature Guide)

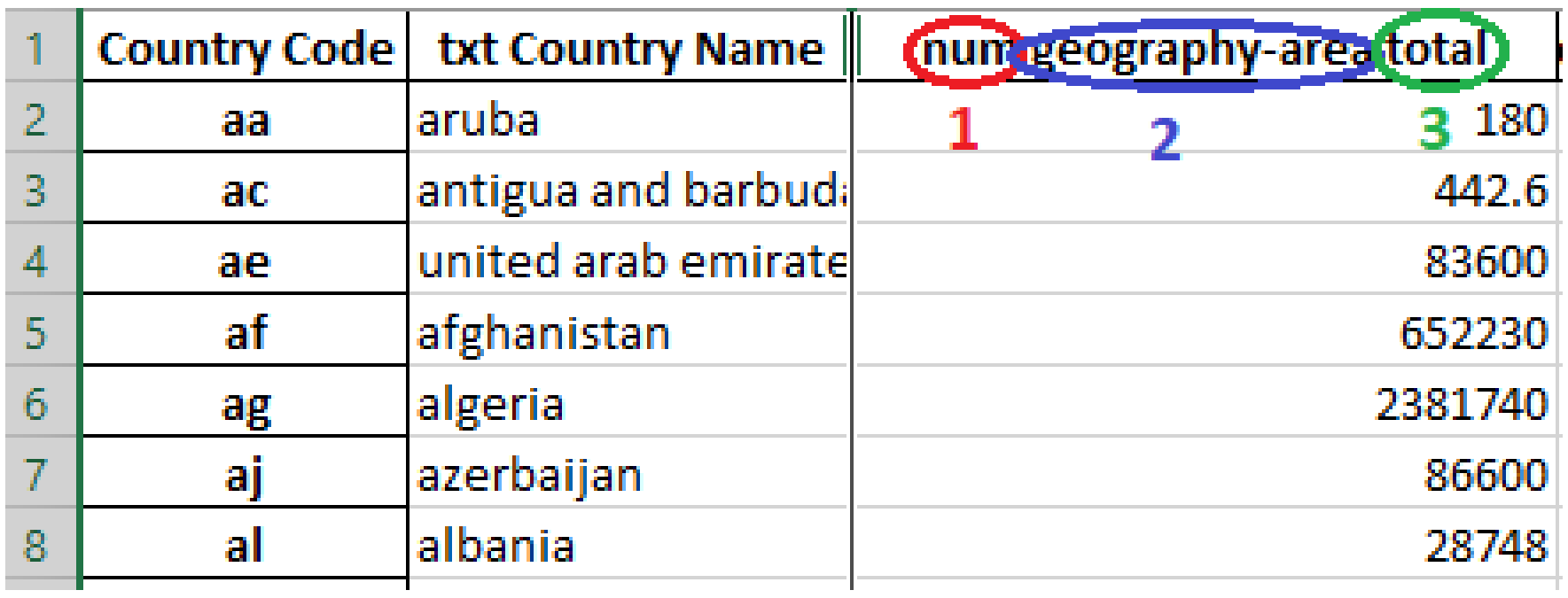

| 1 | Country Code | txt Country Name | num geography-area total |
|---|---|---|---|
| 2 | aa | aruba | 180 |
| 3 | ac | antigua and barbuda | 442.6 |
| 4 | ae | united arab emirate | 83600 |
| 5 | af | afghanistan | 652230 |
| 6 | ag | algeria | 2381740 |
| 7 | aj | azerbaijan | 86600 |
| 8 | al | albania | 28748 |

*Figure 4 – Example of project’s Dataframe column nomenclature.*

Two major problems were encountered in this phase. The first problem was the structure of data referring to terrorist groups which led to numerous columns being created with most of them being more than 90% empty due to the fact that most terrorist groups operate in a small minority of states globally. The second problem stems from the fact that subfields of non-state entries (examples are: “World”, “European Union”, “Antarctica” etc.) are structured in a different manner and single data columns are created. Both problems were tackled on the next phase of the project with cleaning and preprocessing.

The Dataframe resulting from the first phase of the project has:

| 1st version Dataframe (Raw Factbook) | |
|---|---|
| Rows | 268 |
| Columns | 539 |
| Total Size | 144452 cells |
| Empty Cells | 84129 (58%) |
| Data-containing Cells | 60323 (42%) |
| **Values of this table slightly vary compared to Appendix B. Overall Dataframe Versions, Processing, Shape due to different versions of algorithms executed.* | |

*Table 2 - 1st version Dataframe basic characteristics.*

A quick exploration of our dataset as shown in the above table found that 58.3 % of cells in our 2-dimensional table were empty. It became evident at this stage that even though our data are of high quality, our dataset isn't.  Further exploration of our data uncovered that a) The CIA does not offer complete data for each entity, b) the data extraction code has weaknesses due to the inherent lack of structure in the data. Problem a) was solved at a later stage with missing values imputation and missing not at random detection while problem b) had to be tackled at the data-cleaning stage which will be describe in-detail below.

The weaknesses of the dataframe-populating algorithm are two. Firstly, it spots Fields and creates the relative column before moving on to look for contained subfields and the second weakness is its inability to process grouped subfields' data hence appending them as columns. As a consequence, 25 columns only contain a single value, while approximately 40 columns contain data as part of their title. Regarding data structure, the fact that not only states but also overseas territories, geographic regions, the European Union and other non-state entries are included, leads to an increase of data noise and decrease of overall structure and code performance. For this purpose, a data-cleaning script was developed aiming to reduce the noise and data size while preventing data loss.

Despite these weaknesses, the data extraction and preprocessing phase captured, defined and preprocessed all 268 CIA World Factbook Entries and all subfields. Big portion of the Dataframe is empty because of the reasons discussed above. The quality[19] of the Dataframe was greatly improved over the next phases.

## 2.3 Data Processing

### 2.3.1 Data Cleaning (Size Reduction)

The project aims to capture as much CIA World Factbook data as possible and then present them in an easy to interpret and ready-to-use dataset. Unfortunately, the Factbook's uneven and incomplete data collection for each entity and the presence of non-state entities which follow different structures, led to an important number of empty cells (in the Dataframe).

[19] In terms of missing data and shape.

Considering the current state of data science, the output-dataset of this project will surely be used, and has already been used as presented in Chapter 3 – Relevant Projects, for machine learning thus an important problem emerges: The small number of entities (observations) when combined with multiple missing data inside columns (features) render it almost useless for machine learning (see curse of dimensionality). The reduction of empty cells was achieved with the following approaches in the respective order: Row Removal from the dataset, column concatenation, filling of MNAR columns (ex. txt people-and-society-major-infectious-diseases food or waterborne diseases which does not appear in countries with no such diseases and is at first considered as nan while in reality true value is 0 - for all columns assumed MNAR see Appendix D. Columns Assumed MNAR ), column removal, and missing value imputation which is covered in 2.4 Modeling and Evaluation (Missing Values Imputation). Special attention was given in maintaining an equilibrium between data loss and missing values reduction during the data selection and cleaning processes.

The data cleaning process included the following actions in the order presented below:

1. Entries (rows) removal. Entities which do not represent states (ex. world, European union, Indian ocean etc.) were dropped. Such entries increase the dataset's noise by having different structure. Moreover, their role is heavily debated and disputed in the study of International Relations placed mostly by realists outside of the International System or considered as tools of states. The EU can be an exception to the above rule but for the time being and considering that it lacks significant political integration it was removed. Entries which do not have an official population figure were also dropped. Most of such entries were overseas territories and tiny islands. Once again, such entries are exceptions to the rule (in terms of demographic, geographic and political characteristics) with outlier-like characteristics and do not seriously affect the international system on their own.
2. Column concatenation: In the first version of the dataset, every terrorist group was placed as an individual column. All terrorist groups were concatenated into a new column which provides a list of terrorist groups per country. Similar was the process applied to the bordering countries columns which were also concatenated into one. This process alone, reduced empty cells by 40,2% from the 1st version Dataframe.

3. MNAR filling. All columns were manually reviewed by the author, some were assumed to be Missing Not At Random (MNAR) and filled with zero values (where missing). For a list of columns assumed MNAR see Appendix D. Columns Assumed MNAR.
4. Column removal. Columns with more than 95% of their data missing were removed. A lower threshold was not selected to avoid removing MNAR columns which were not identified successfully in step 3 and may be discovered in future studies of the dataset.

After the execution of all data-cleaning steps, the second version[20] of the dataframe, reduced by 42 rows and 224 columns was created. The new Dataframe was shaped: 226 rows x 315 columns.

The size reduction/cleaning algorithm managed to decrease empty cells by 87% while losing only 1.36% of data[21]. The first version Dataframe had: 144452 total cells, 84129 empty cells, 60323 data-containing cells and the second version: 71190 total cells, 11081 empty cells, 60109 data containing cells. Moreover, the loss of information becomes even less significant if the qualitative characteristics of our data are taken into consideration. Existing columns of textual data can help us discover and generate more data.

### 2.3.2 Data Construction and Encoding

Our dataset has an important amount of textual data which contain valuable information. This information needs to be extracted not only because it is important for the study of IR phenomena but also in order to be used as machine learning features. In every case, textual data need to be converted into numerical. After studying the dataset, a non-exhaustive list of columns was formed by handpicking columns. This selection was done on the basis of data importance and structure. Time constraints led to a sample of columns being chosen for further processing. In that regard, 39 columns were selected for further processing.

[20] For a summary of Dataframe Version see Appendix B.
[21] Rough figures which cannot properly calculate the qualitative characteristics of data.

The 39 columns were examined before determining that data can be mined in 3 ways (methods):

1) Data can be converted into categorical/labels (ex. Dependency Status, Tier ratings, Government type etc.)
2) Amounts of elements can be calculated (ex. Amount of terrorist groups, amount of ports, amount of natural resources, amount of merchant ships etc.)
3) Sum of elements can be calculated (ex. Sum of refugees, length of pipelines etc.)

Some columns can be broken down to more than one of the above. For example, column "txt transportation-ports-and-terminals container port(s) (teus)" contains both the number of container ports and the volume of cargo transferred through each one.

Columns were then divided into 4 groups depending on which of the above methods can be applied to them. Group 1 included columns ready to be converted to categorical, Group 2 contained columns out of which only amounts could be calculated, Group 3 was formed by columns out of which only sums could be calculated and lastly Group 4 consisting of columns which needed special (per-column) processing and were subject to any combination of the aforementioned methods.

Similar code was prepared and executed for Groups 1,2,3. Specifically, code made use of regular expressions and of basic python operations in order to deconstruct text and provide with the desired outputs.

For columns of Group 4, different code was developed for each column. This process was the longest in terms of both time consumed and code size. The highly unstructured nature of data raised serious challenges not only in terms of code but also in terms of data distortion. Each and every column of Group 4 was studied and relative structure of data was determined[22]. In many instances, data of the same column had different structure, probably due to the primary data-collecting agency and not the CIA. After a column was studied, the potential data conversions were determined. The data were then deconstructed and manipulated with attention to avoid losing data or distorting them.

[22] As an example, in many cases like column txt government-suffrage (for Argentina) : *"18-70 years of age; universal and compulsory; 16-17 years of age - optional for national elections"* the symbol ";" divided the general data element from information. The dividing symbol was then used to split the sentence and the left part was used for further processing.

Unfortunately, in one instance, due to the lack of standardization, data had to be supplemented from 3rd party sources. This was the case for: Column "txt geography-climate" (example of data for the UK: *"temperate; moderated by prevailing southwest winds over the North Atlantic Current; more than one-half of the days are overcast"* (CIA, 2019)). It was determined that the only possible output is categorical/label. Sentences were searched for punctuation characters (; : ,) in the respective order. Sentences were then split at the highest-level symbol found and the left-most section was picked for further processing. The string segment resulting from the previous splitting operation was searched for keywords (tropical, semiarid, temperate, etc.). Keywords were selected after reviewing data for common occurring words. Depending on the keyword found, each country was assigned a climate type based on the Köppen climate classification system (Beck, et al., 2018). For more information on conversion of textual columns see Appendix E. Columns Generated from Textual Assumptions.

New columns created from the Processing phase (source columns were not replaced) were named using the original title (ex. txt military-and-security-military-branches) but with the data type indicator (in this case txt) replaced either by "lbl" for categorical data (ex. lbl government-citizenship citizenship by birth), "sum" for summed amounts (ex. sum transportation-pipelines, which originally included lengths of various pipelines) or "amount" for amount of original elements (ex. amount military-and-security-military-branches). Empty cells of category "lbl" were filled with string "None/NA" and empty cells of category "num" or "amount" were filled with number 0 if believed to be missing not at random (MNAR). For example, developed countries don't have major infectious diseases or home-based terrorist groups but data appeared as missing. This was fixed by placing 0 or nan in place depending on data type. The third version of the dataset/Dataframe was introduced at this point (For a summary of all datasets/Dataframes see Appendix B). All new columns were then appended to the second version of the Dataframe giving birth to the third version dataframe with 52 new data-complete columns.

Lastly, categorical/label "lbl" columns were converted into dummy/indicator ones using the pandas' "get_dummies" (Wes McKinney and the Pandas Development Team, 2020, p. 963) method which in reality is One-Hot Encoding.  The new columns generated were numerical (1/0 binary) while the original columns were textual-categorical. For the new dummy-columns "enc" was used as data type indicator. These dummy columns were then appended to the dataset/Dataframe creating the $4^{th}$ version which is the largest version with 227 rows, 500 columns (column types and amounts: 'lbl': 16, 'amount': 18, 'sum': 18, 'txt': 122, 'num': 192, 'enc': 134). All columns of type amount, sum, enc and few num columns were complete in terms of data and were used as features for the estimation of all missing numerical values.

In the case of sanitation and water source columns, the provided values refer to the unimproved conditions. In cases where ranges of values were provided (ex. refugees 2.5-3.0), the largest value was chosen.

Overall, it should be noted that columns produced from this phase of the project (all those having "lbl", "amount", "sum", "enc" as data type indicator) are not always expected to be 100% accurate and attention should be given when used.

## 2.4 Modeling and Evaluation (Missing Values Imputation)

### 2.4.1 Model Selection, Evaluation and Design

This project suffers from one of the most common problems in Data Science, missing values, which contributes negatively to a second problem, dimensionality. Missing values do not allow for wholistic research to be conducted and seriously limits machine learning capabilities. The Processing algorithm described above, filled only a small portion of missing data. Another option would be to generalize and fill all missing values with 0, mean values or nan but such a choice would seriously hurt the quality and credibility of the dataset. In that context, it was decided that more missing values could be found through machine learning. Specifically, numerical values. The decision to avoid textual missing-data computation was taken on technical grounds. The objective of the models implemented in this project was to successfully fill missing values strictly within the scope of this project and not to be deployed in the real world.

For the selection of the appropriate machine learning models, the following were carefully considered:

1. The author has limited machine learning knowledge and computing resources rendering deep learning models unavailable as an option. Moreover, models selected need to be proven and well-established requiring as less hyperparameter tuning as possible.
2. The dataset consists of textual, numeric and labeled data. The scope of this paper and with the unstructured nature of textual data contained led to the decision that only numeric data will be predicted. The prediction of numeric values led to regression models.
3. Data have a high variance, high dispersion, uneven and differing distributions with many instances of outliers due to variability. There are also instances of multiple groups of columns referring to the same category or field increasing multicollinearity. Random Forests are known to handle chaotic data and many features well.
4. Dataframe Features (columns) outnumber observations (rows) which are relatively few for machine learning. Only complete columns can be used as features in order to achieve minimum observation-loss thus maximum training and testing data.
5. Building a model tailored to every single column is not viable in terms of time. Selected methods were implemented on all columns and the one with the highest accuracy was used for predicting missing values.

The above factors led to Ordinary Least Squares (OLS) Simple Linear Regression, Multiple Linear Ridge Regression and Random Forest Regression models being used. The library sci-kit learn (scikit-learn developers, 2020) was used for all models.

Simple linear regression was chosen because it is fast, simple and efficient reducing the workload of the other models.

Multiple Linear regression because it is simple, fast, efficient and proven but also because it takes under consideration features that the simple regression doesn’t.

Random Forests for being known to handle chaotic data and provide good results with limited hyperparameter tuning.

After having established generic model types, it was crucial that the best model of each category was selected. This applied mostly to linear regression models for which sklearn offers a plethora of models. Not only models but also feature selection methods were compared.

For the comparison of models and methods, it was decided that the Mean Average Percentage Error (MAPE) will be used. This metric not only is easy to use and interpret but is known to be a good option for model comparisons. In order to tackle its biggest weakness, actual 0 values were removed along with the paired prediction in order to avoid divisions by zero. The validity of the metric was increased by the repetitions of the process through K-folds validations which were themselves placed within automated feature selection processes, also executed multiple times. The different ways of splitting the same group of data along with the multiple different iterations raise the confidence in the metric.

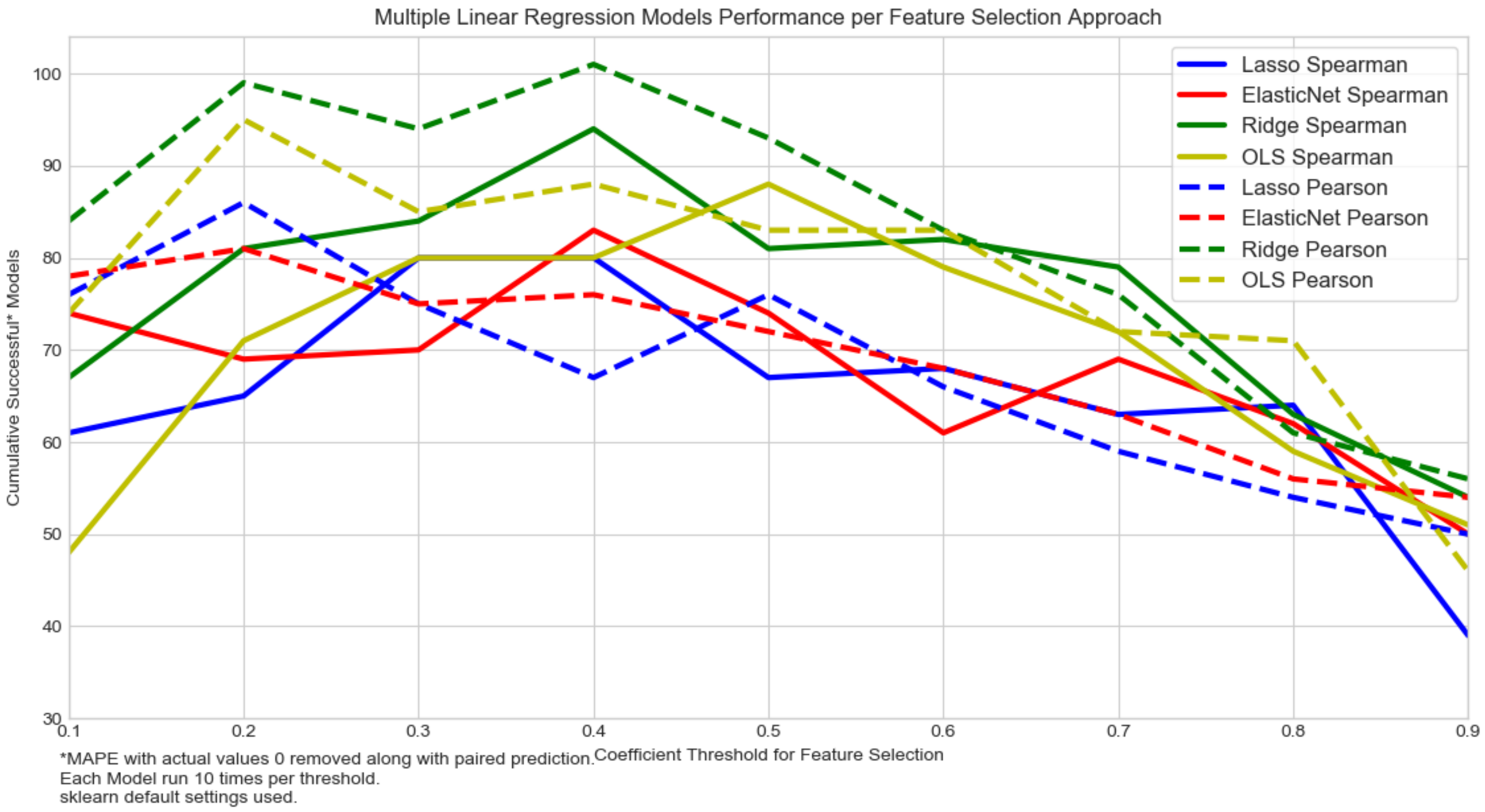

*Figure 5 - Multiple Linear Regression Models benchmarks.*

Each model was run 10 times per feature selection threshold over the whole dataset. Successful models were calculated cumulatively using a modified Mean Average Percentage Error which dropped actual 0 values along with paired predictions. Successful models were considered those with a MAPE smaller than 15%. At the same time, optimal feature selection techniques and thresholds were calculated. Specifically, features were selected based on Pearson's or Spearman's correlation coefficients at values illustrated in the graph above. Both Pearson's and Spearman's coefficients take values from -1 to 1 were converted to absolute values (0 to 1) since strength and not direction of correlation is important. In our study, the only case in which the correlation coefficient equaled to 1 was between a column and itself. Due to that, 9 absolute-value discrete thresholds could be established at the single decimal digit level. Pearson's correlation as basis for feature selection outperforms Spearman's. Contrary to other machine learning projects, overfitting is not a big problem considering that models are not to be used outside of this project. The test conducted above not only serves to identify which correlation type is better for our case but also helps discover the optimal threshold in choosing columns as features. Threshold 0.4 provided the most successful models. The last step in choosing the best Multiple Linear Regression Model required a comparison between the models which select features on the basis of correlation with the model which uses all complete columns as features. Each model was run 10 distinct times for each threshold and the cumulative number of successful models per threshold was calculated. It was found that in between Multiple Linear Regression models, Ridge regression is the most efficient with a Spearman's correlation coefficient threshold at 0.4 for feature selection. Models were then deployed in order with respect to their time consumption:

Single Linear Regression → Multiple Ridge Regression → Random Forest Regression

### 2.4.2 Single Linear Regression

Ordinary Least Squares Linear Regression was selected. This model was used on the data before all other models. Being the simplest in terms of feature selection and hyperparameter tuning, it can easily compute missing values using only one feature. Bivariate[23] Pearson's Correlation coefficients were calculated between the target column (label) and all non-missing data columns. The complete column with the highest Pearson's[24] coefficient was selected as feature. Data were then split into training and testing at a 80-20 ratio. Parameters were set automatically with 5-folds Grid Search cross-validation scored by MAPE[25]. Every model with a MAPE <15% was considered successful and implemented on the data. Correlations were then recalculated to include any newly completed columns.

The model was run 10 times per column missing data[26] and performed:

| Single Linear Regression Model Results | |
|---|---|
| Successful models/columns completed | 39 |
| Cells filled | 1451 |
| Missing Value Reduction | 18.2% |
| Successful models considered those with MAPE <15% | |

*Table 3 - Single Linear Regression Model Results*

Even though Single Linear Regression is concise in terms of outcomes, the model was run 10 times per column because of the Grid Search Cross Validation which can provide different hypermeters due to the randomness in splitting the data for training, testing and validating. Thus, every repletion of the model can lead to the creation of more efficient models not built in previous iterations.

---

[23] (between the target column and feature column)
[24] The Pearson's correlation coefficient describes linear correlation.
[25] In the model evaluation phase, actual values of 0 were removed along with the predicted value in an effort to avoid nan/inf MAPE outcomes.
[26] More iterations would have yielded results of higher confidence but computing resources did not allow for that.

### 2.4.3 Multiple Linear Regression

Ridge Multiple Linear Regression was performed after consulting the findings presented in Table 4 at 2.4.1 Model Selection, Evaluation and Design. The same table helped identify the optimal feature selection method being the Pearson's correlation coefficient. Structure-wise, the algorithm was identical to the Single Linear Regression one, with the only exception being that not one but rather a group of columns were used as features after considering correlation thresholds.

| Ridge Multiple Linear Regression Model Results | |
|---|---|
| Successful models/columns completed | 11 |
| Cells filled | 282 |
| Missing Value Reduction | 3.5% |
| Successful models considered those with MAPE <15% | |

*Table 4 - Ridge Multiple Linear Regression Results*

### 2.4.4 Random Forest Regression

For the Random Forest Regression, scikit-learn's RandomForestRegressor (scikit-learn developers, 2020) was used. Contrary to the Linear Regression methods discussed above, the Random Forest Regression can better capture the complex and varied relations of our data. Unfortunately, a higher degree of hyperparameter tuning is required. Grid search cross validation was compared with the baseline model (hypermeters at default) and, interestingly, performed worse. Specifically, after 10 executions of each code, the Grid Search cross validated approach provide with 2 successful models over a 2-hour period while the baseline model provided with 6 successful models in 2.57 minutes.

| Random Forest Regression Models Comparison | | | |
|---|---|---|---|
| Approach | Hypermeters | Feature Selection | Successful* Models |
| A. | Default | All Columns | 0 |
| B. | Default | Importance Weights | 6 |
| C. | Grid Search Cross Validated | All Columns | 2 |
| Successful models considered those with MAPE <15% | | | |

*Table 5 - Random Forest Regression Algorithms Comparison*

Approach B was implemented using scikit-learn (scikit-learn developers, 2020) "sklearn.feature_selection.SelectFromModel". Importance weights were drawn from the Random Forest Regressor itself. Train and test data were transformed separately and after splitting in order to avoid contamination or bias. The implementation of the algorithm yielded the results described below. It should be noted that the data had already been subject to the Single Linear and Multiple Linear Regressions described above with sixty columns already filled.

| Random Forest Regression Model Results | |
|---|---|
| Successful models/columns completed | 6 |
| Cells filled | 285 |
| Missing Value Reduction | 3.5% |
| Successful models considered those with MAPE <15% | |

*Table 6 - Random Forest Regression Results*

## 2.5 Strengths, Weaknesses, Future Ideas

The present paper covers an underrepresented topic in IR by increasing the existing limited literature. It hopes to become a breakthrough by further introducing the ideas of "Computational IR", Data Science and of the "Fourth Paradigm". Thus, it offers a new perspective on the field. It challenges and invites researchers and practitioners to experiment with this new approach in a wholistic manner. Moreover, it constitutes an anthology of basic tools and approaches ideal for IR students who want to explore the world of Data Science within the scope of their studies.

The backbone methodology (CRISP-DM) is one of the most if not the most widely acceptable and proven in various fields of study and applications.

Hard to dispute data of high quality are drawn from the prestigious US Central Intelligence Agency and its decades-old and widely recognized encyclopedia "World Factbook".

The code developed for this project is modular, open-source and can be applied both to future and past publications of the Factbook.

There is a great window of potential for further development of both the ideas and code of this paper.

The reporting of findings in Chapter 3 – Relevant Projects along with the provision of a ready-to-use dataset all support the main argument that Data Science can be extremely useful for the IR scholar and decision-taker. This is a rare case in IR of a theoretical argument coexisting in an otherwise rather technical paper. This coexistence not only serves to support the argument but also to enable them conduct further research or built upon this very project themselves. This effort brings the contemporary ideals of the internet in the study and practice of International Relations by employing a collection of computational resources.

Prior to analyzing technical weaknesses, it is crucial that some theoretical warnings are raised. It should be noted that data science is no panacea nor totally supported by axioms, especially in social sciences. It merely constitutes a major tool supplementing the field. In that spirit, Campbell's L and Goodhart's Laws should be reminded: *"The more any quantitative social indicator is used for social decision-making, the more subject it will be to corruption pressures and the more apt it will be to distort and corrupt the social processes it is intended to monitor."* (Campbell, 1979) and *"When a measure becomes a target, it ceases to be a good* *measure."* (Strathern, 1997). The International Relations data researcher should never assume a priori that reality can be totally explained and forecasted with models nor completely disregard existing literature and history for the sake of data.

On a technical level, the largest weakness of this project lies in the missing values imputation phase. This phase is very reliant on Mean Average Percentage Error which is calculated without using actual 0 values and paired predictions in order to avoid division by zero. As a result, models are evaluated, selected and implemented based on a single metric whose validity is still widely debated (as happens with most of the other related metrics).

The creation of new numerical columns derived from textual ones has inherent weaknesses. The complexity of written language and the lack of uniformity and structure does not guarantee 100% accurate data transformation.

Overall, the multifaceted character of this project makes its structure, presentation and wholistic comprehension demanding tasks.

The author plans to optimize the code in the future ultimately aiming towards a totally object-oriented program. The Factbook will constitute the object and the user will be able to perform various operations breaking the original pipeline of the project. Moreover, such a program could be made a library specifically oriented towards IR students and researchers. In the latter case, a user guide may also be produced. For the time being, all aspects of the code can be further enhanced with the most important and time demanding being the Machine-learning Missing value imputation phase which could be replaced with deep learning models to provide more tailored predictions. The missing values imputation process can be also strengthened if more metrics[27] are used to better evaluate models. There is also large room for improvement when it comes to feature selection. In no case has the author exhausted or tested all possible approaches.

NLP tasks (Liddy, 2001) can help mine large quantities of data from the textual columns which have seen limited use in relevant projects. Vectorization could provide with a smoother transformation of textual columns to numerical but at the cost of losing human interpretability.

[27] Like MSE, RMSE, MAE etc.

The generated dataset can be used in every type of machine learning project. It can be combined with other data in order to provide forecasting models or even be expanded to include older Factbook Data (2000-2018 data available) in order to produce timeseries forecasting models or even assist in visualizing the effects of country policies within the last two decades.

PCA can provide with strong indexes for each country embodying a great wealth of information per sector (ex. Transportation, Energy etc.).

Lastly, this dataset can be used for classification tasks like developed-developing countries which can then be used in other projects as labels.

MNAR columns will be revised and further studied.

The potential of this dataset is only limited by the imagination and knowledge of the user. From a statistical point of view, it also provides with data for the whole population of countries and not samples. This dataset can be best utilized in the context of working groups consisting of members with technical backgrounds (mostly statistics, algebra and programming) with close consultation with International Relations researchers who need to possess an adequate technical understanding.

The code and the datasets of this project will be shared online after the publication of this paper and will keep getting updates on the long run with the input of the online community.

# Chapter 3 – Relevant Projects

In this chapter, a collection of various applications and short projects making use of the produced dataset are presented. This chapter aims to provide with ideas and examples of how this dataset can be used in practice with tangible examples and applications. All of the projects were created by the author of this paper for various undergraduate courses.

## 3.1 CIA World Factbook Sentiment Analysis

In this application, all textual data of the CIA World Factbook were analyzed with statistical and Natural Language Processing techniques and algorithms in order to determine whether the CIA World Factbook, and the US as an extension, portray allied states in a more favorable (positive) manner. In doing so, the Factbook dataset was successfully and easily combined with a different dataset (opinion poll by YouGov). The research conducted concluded that: *"From 1,2 and 3 it was found that there is a weak correlation between the policy and sentiment of Friendly, Allied and Unfriendly states. Friendly and Allied states are portrayed more favorably than Unfriendly".* Lastly, various sentiment analysis algorithms were compared.

The approach and ideas of this paper can be used by a government to reverse engineer their true relationship with the US. Furthermore, it also serves as verification of the findings of public opinion polls (like the YouGov one used) in an effort to create bilateral relations labels between the US and other countries. Lastly, it can also be used as a STRATCOM tool to promote a certain narrative either in favor or against a bilateral relation with the US.

## 3.2 Religion in International Relations

This work was a poster prepared for the undergraduate course "Religion and International Politics". It aimed to provide with a clear and inclusive image of the global religious scene. It was an approach carried out with data visualizations, K-Means and MeanShift clustering implemented with scikit-learn (scikit-learn developers, 2020). All of the above were carried out only with the use of the CIA World Factbook dataset. At first, the column referring to the regime of countries was converted from textual to categorical and the column describing religions of each country was transformed to multiple numerical ones. Then, religions per regime type were visualized to discover that:

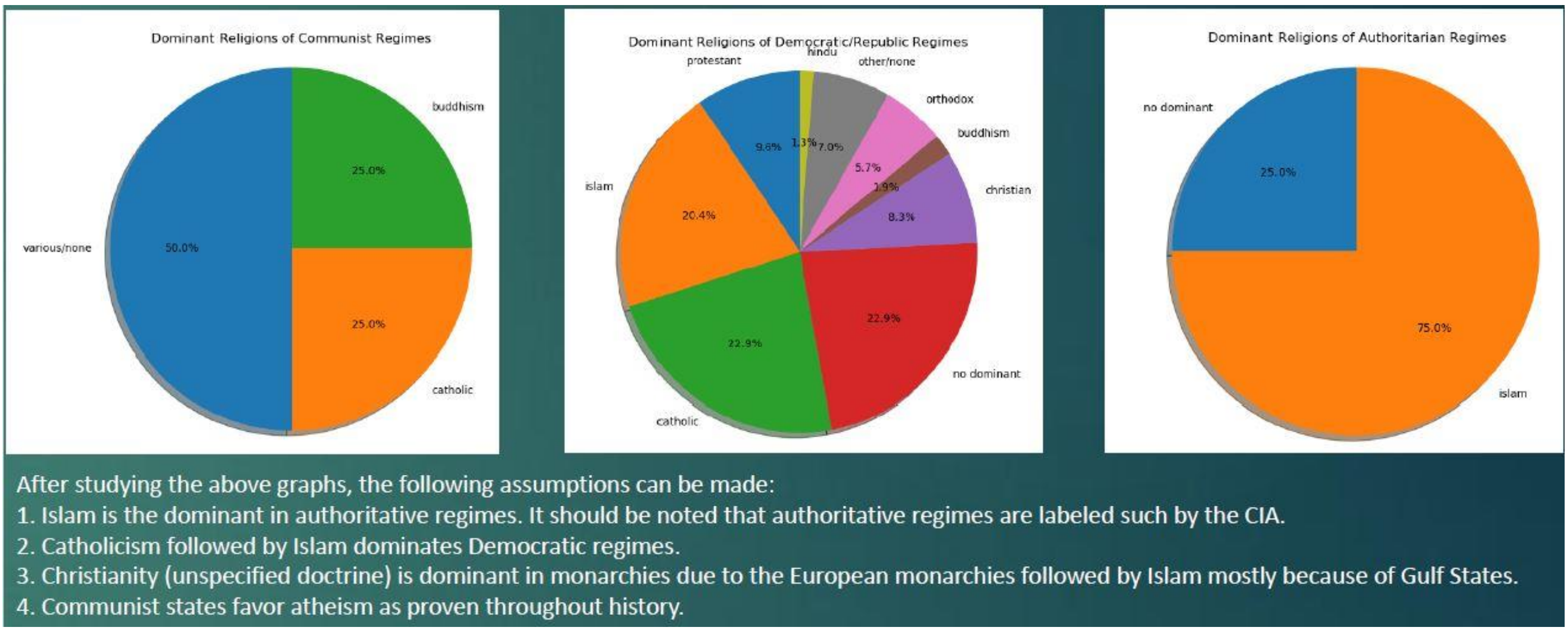

*Figure 6 - Partial Findings of "Religion in International Relations Paper"*

The Clustering of all CIA World Factbook Countries and the descriptive statistics (Eric, Travis, Pearu, & others, 2001) of each cluster revealed several more interesting findings all of which are difficult to summarize in this paper. A selection of some is: *"Islamic religions seem to be the strongest dominant religions (higher average % of any other dominant religion); meaning that Islam is a stronger dominant religion than other religions. Furthermore, countries with percentages of Islam followers tend to have less religious diversity and atheism. Protestantism seems to occur in countries with the highest religious diversity".* Catholicism seems to be stronger than Orthodoxy and Protestantism in terms of dominant religion. Moreover, Protestantism seems to be the "softest" dominant religion. Israel and Armenia (which both have unique dominant religions) were clustered along atheist and non-dominant religion states with the K-Means[28] algorithm and each one alone with the MeanShift[29] clustering meaning that these two countries are exceptions to the rule and should be examined on a per case basis.

The above are only a fraction of findings from the related paper (poster). Those interested are strongly advised to read it. The usefulness of the aforementioned paper lies mostly on the grounds of education, research and advising as it can help convey large amounts of information regarding countries and religions with less than 1000 words while fueling future research and hypotheses.

## 3.3 The Turkey/Qatar – Egypt/UAE/Saudi Arabia blocs

This project compared the two blocs of countries mentioned in the title. After having understood and presented the political background, a data-centered comparison was conducted using 227 columns of data from the CIA World Factbook. Saudi Arabia – United Arab Emirates and Egypt were found to be a more homogenous bloc than Turkey – Qatar: *"The biggest discrepancies between the states of bloc 1 (Turkey-Qatar) refer to a wide array of sectors including energy (gas production, LNG terminal, pipelines, oil imports/exports), geography (irrigated land, forest land), transportation (highways, ports, airways), climate, government etc. At the same time, fewer and smaller differences in bloc 2 (Saudi Arabia – UAE-Egypt) are associated with indexes of lesser importance (ex. diseases, naturalization age limits etc.)".* Moreover, exact differences and similarities between these two blocs were calculated presenting the strengths, weaknesses and complementarity of the two blocs; an important finding for decision-takers.

Finally, after conducting clustering on the five countries and combining all findings of the paper with the political and IR background, it was concluded that: *"It is once again proved that interest is the dominant factor in alliance forging... Moreover, considering that countries of blocs 1 and 2 tend to cooperate with states with similar forms of government may also suggest that countries with same government types are more eager to cooperate but further analysis is suggested on the matter...political factors remain strong compared to economic ones*

[28] Supervised method working with Euclidean distances, number of clusters is provided by the programmer and determined mostly with the elbow and silhouette methods.
[29] Density based - Unsupervised method which determines the number of clusters without human interference.

*due to the fact that even though Qatar has more intense financial transactions with neighboring Gulf states it chooses to side with Turkey mostly because of political matters…both Tukey/Qatar and UAE/Saudi Arabia/Egypt blocs are short-term tools and not long-term agents of change…Both strong armies and economies (mostly for financing others as in the case of Egypt - Saudi Arabia) are vital elements of regional politics."*

## 3.4 Global Energy Dynamics

This application visualized energy related CIA World Factbook data on a regional (continental) basis and then performed various clustering[30] implementations regarding individual countries. The visualizations offer an easily interpretable idea of the global quantitative dynamics. Scaled versions of the graphs also reveal qualitative characteristics between continents. Clustering and descriptive statistics helped form the following (among other) conclusions:

1. *"The world consists of the following region groups in regards to energy:*
    a. *Group A (Central America, S. Asia, Oceania, S. America, N. America): Regions which are relatively autonomous. They produce decent amounts of resources without being considered powerhouses. Equilibrium of imports/exports. Probably the most energy secure and stable regions.*
    b. *Group B (Middle East): Focused on crude oil, still has strong natural gas potential. Biggest reserves, very high production which can be further increased. Lack of internal trade, many global exports. The highest potential for political influencing (already a reality) but also for long term resource production.*
    c. *Group C (Central Asia): The Middle East of natural gas. Massive natural gas reserves and medium crude oil ones. Very high production of gas and high production of oil. An unsustainable but strong short-term development strategy. Many exports, mostly towards Europe (Russian gas). Need for investment in non-energy sectors. Short term development at the expense of sustainability.*

[30] Mean-Shift clustering on scaled data: Density based - Unsupervised method which determines the number of clusters without human interference.

*d. Group D (Europe): Diversified electricity production, biggest importer of energy resources. Minimal reserves and production except refined petroleum products which mostly come from the manufacturing sector and are circulated internally. Global commercial hub with extensive infrastructure (pipelines, ports etc.). High cost of energy and low security. The EU is aware of all these and is increasing investment and reliance on renewable sources.*

*e. Group E (East/Southeast Asia): Similar profile to Europe but without exports. Probably the most energy dependent region of the world, combined with relatively low living conditions and lack of regional trade. A track similar to this of Europe would be ideal. Diversification could start with hydroelectric which is cheaper than renewable.*

*f. Africa*: Africa proved to be the most versatile region being clustered differently every time new features were added. It presents characteristics from both Group A and Group D but at the same time important differences with both. It has a decent production of resources like Group A but very active imports and exports. At the same time, its intense trade looks similar to this of Europe but there is a great differentiation in the selection of electricity production sources with Africa being extremely fossil-fuel reliant. Very high potential for investment and development. Economies with low demands, central geographic location and variety of countries.*

*2. The clustering of countries distinguished Canada, China, France, Germany, Italy, India, Iran, Japan, Netherlands, Qatar, Russia, Saudi Arabia, Singapore and Venezuela. These countries constitute interesting case studies. “*

This paper is not only useful within university amphitheaters but can also assists real world policies and private sector investment.

## 3.5 Education and International Relations

As the title suggests, this paper focused on education. Contrary to other projects presented in this section, this one focused more on education itself rather than on international relations. It made use of CIA World Factbook, UN and OECD data (combined) in order to find the correlations between the quality of education in a country and more than 100 other factors. After using various tools, it was ultimately found that:

*"1. While commenting on the bivariate correlations, the developed versus developing country dichotomy kept repeating. Further research on whether education can be a key feature for the classification of countries into developed and developing ones is suggested. While the correlations revolved around the developed – developing dichotomy, clustering did not seem to prove it.*

*2. There is a general lack of education-related qualitative metrics hence the failure of clustering algorithms. Can the quality of education be calculated indirectly through strongly correlated data?*

*3. European countries offer the highest quality education and African states the lowest, homogeneity considered.*

*4. Education is a complex multi-correlated phenomenon. Closely related with life expectancy and sanitation. Lower and fewer correlations with financial factors.*

*5. Participation of women in education is a strong measure of educational system quality.*

*6. The clustering algorithm showed inability to adequately group countries. This leads to the assumption that countries either have diverse and unique educational characteristics which could be better studied on a "per case" basis or that existing measures (like School Life Expectancy and UN's, PISA scores) are unable to provide with an accurate depiction of educational realities.*

*7. More attention should be given to the correlations section of this paper and not on clustering.*

*8. Healthcare, sanitation and social factors are more correlated with education than economic ones. "*

This paper provides ample information and ideas regarding not only the future study/research of education but also for the creation of more efficient education policies. It successfully managed to draw data from IR and by using data science and statistical methods and tools, it generated insights which can be used in a variety of sectors spanning from social science research to gender studies and governance.

***USER GUIDE NOTICE!**

Under no circumstances can this section be considered as a complete User Guide. The author is planning to produce a user guide over the next year after having received and implemented community and academic feedback on this project.

# Appendices – User Guide*

## Appendix A. Program Pipeline

*DF: Dataframe (for the detailed list of dataframes see Appendix B. Overall Dataframe Versions, Processing, Shape)

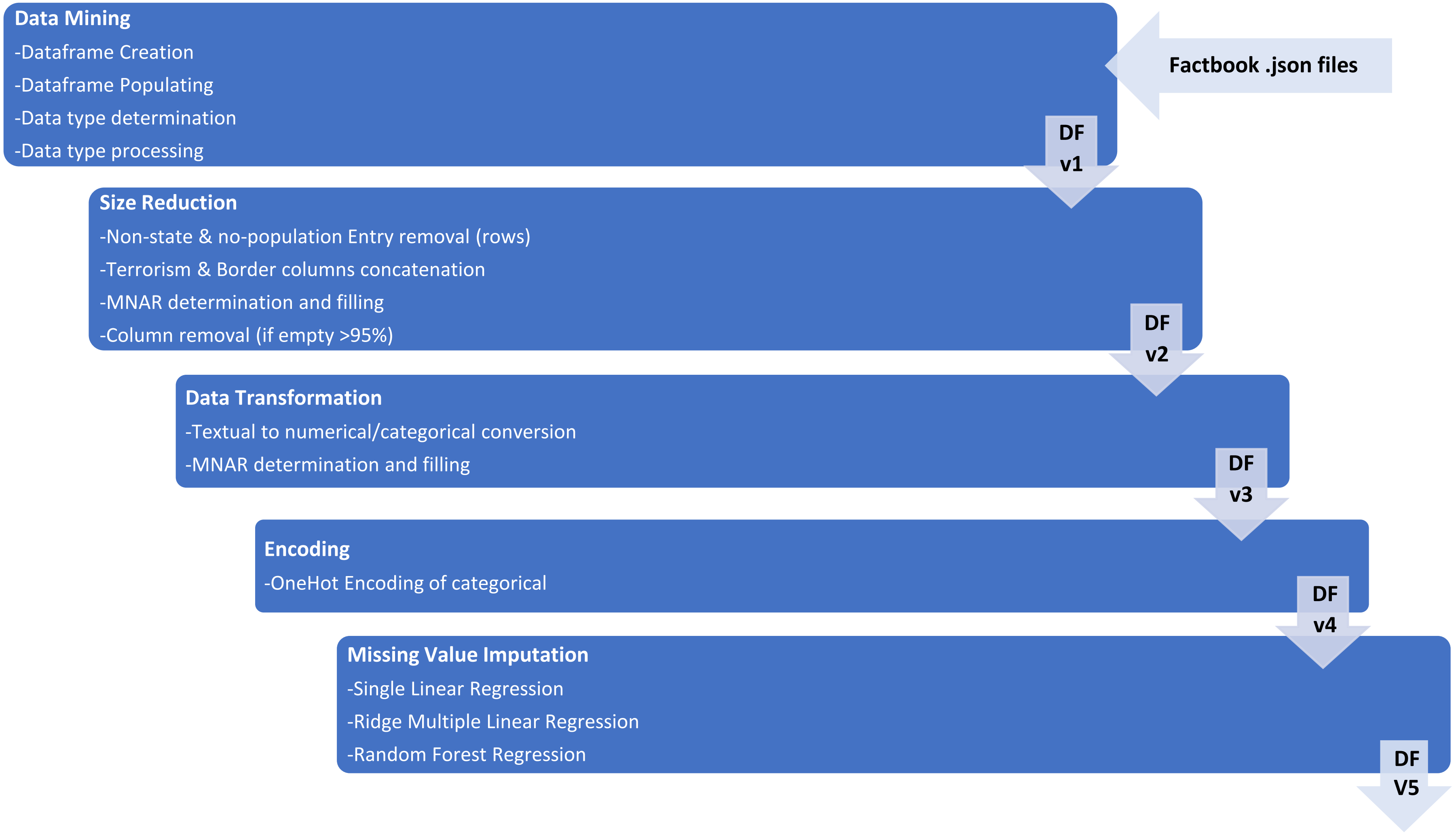

## Appendix B. Overall Dataframe Versions, Processing, Shape

| Version | Status Description | Rows, Columns | Empty Cells | Data Containing Cells | Columns Data Types |
|---|---|---|---|---|---|
| v1 | Data Extracted (data types found and processed) | (268, 540) | 85513 | 59207 | Labeled: 1,<br>Textual: 314<br>Numerical: 225 |
| v2 | Size Reduced (rows, and columns removed or concatenated, MNAR filled) | (237, 322) | 7043 | 69271 | Labeled: 1,<br>Textual:134<br>Numerical: 187 |
| v3 | Textual converted to categorical and/or numerical, MNAR filled | (237, 377) | 7949 | 81400 | Labeled: 18<br>Textual: 134<br>Numerical: 225 |
| v4 | Categorical encoded to binary with One Hot Encoding | (237, 500) | 7952 | 110548 | Labeled: 18<br>Textual: 134<br>Numerical: 225<br>Binary Encoded: 123 |
| v5 | Missing values imputed with Machine Learning | (237, 500) | 6385 | 112115 | Labeled: 18<br>Textual: 134<br>Numerical: 225<br>Binary Encoded: 123 |

## Appendix C. Column Nomenclature Guide

| | |
|---|---|
| <u>Column naming format:</u> | [data type]_(“MAPE”):*_[number]*_[category]-[field]_[subfield]*_”hist”* |

**<u>Naming Format Symbol Legend</u>**

| Symbol | Meaning |
|---|---|
| * | optional |
| _ | 1 gap character |
| hist | data available for multiple years, the latest was picked |
| (MAPE): [number] | Column had missing data which were calculated with ML along with the Mean average percentage error of the model. |

**<u>Notes</u>**

Characters “-“ and “ “ do not necessarily always divide categories-fields-subfields. In some cases they appear within those. Specifically, some categories may contain “-“ themselves like people-and-society, not to be considered dividers between category and field. The same applies to some fields as well (example: labor-force-by-occupation which is a single field). Subfields may contain “ “ in between.

| <u>Examples</u> | <u>Explanations</u> |
|---|---|
| num (MAPE): 1.05 economy-gross-national-saving hist | “num” denotes that numerical data are contained. “(MAPE): “ means that missing values were calculated using machine learning. “1.05” is the Mean Average Percentage error of the model which imputed missing values. “economy” is the category of data. “gross-national-saving” is the field. “hist” means that data for various years were available, the latest were picked. |
| enc people-and-society-major-infectious-diseases degree of risk_high | “enc” means that contained data are binary numbers (1 = yes and 0 = no). “people-and-society” is the column’s category, “major-infectious-diseases” is the field, “degree of risk” is the subfield, “_high” is the label. Countries with 1 in this column have a high degree of risk regarding infectious diseases. |

**<u>Data Types Legend</u>**

| Text | Data type |
|---|---|
| txt | Textual data |
| lbl | Text labeled data (ex. Africa/Europe etc.) |
| enc | Binary encoded data (ex. Country in Africa? 0 = country not in Africa) |
| num | Numerical data |
| sum | In essence numerical data but generated manually from textual columns by summing amounts (ex. volume of cargo passing through ports) |
| amount | In essence numerical data but generated manually from textual columns by counting items (ex. number of ports) |

## Appendix D. Columns Assumed MNAR

| Columns Assumed to be MNAR and filled with 0 where values missing |
| --- |
| Number of columns: 105 (without counting columns generated from those) |
| txt geography-land-boundaries-border-countries-overall, num geography-land-boundaries total, num geography-maritime-claims territorial sea, num geography-maritime-claims exclusive economic zone, num geography-maritime-claims contiguous zone, num geography-maritime-claims continental shelf, num geography-maritime-claims exclusive fishing zone, num geography-land-boundaries total, txt people-and-society-major-infectious-diseases aerosolized dust or soil contact diseases, txt people-and-society-major-infectious-diseases soil contact diseases, txt people-and-society-major-infectious-diseases respiratory diseases, txt people-and-society-major-infectious-diseases animal contact diseases, txt people-and-society-major-infectious-diseases water contact diseases, txt people-and-society-major-infectious-diseases degree of risk, num people-and-society-children-under-the-age-of-5-years-underweight, txt people-and-society-major-infectious-diseases vectorborne diseases, txt people-and-society-major-infectious-diseases food or waterborne diseases, num people-and-society-hiv-aids-deaths, txt people-and-society-people-note, txt government-dependency-status, txt government-dependent-areas, txt government-diplomatic-representation-in-the-us chief of mission, txt government-diplomatic-representation-in-the-us chancery, txt government-diplomatic-representation-in-the-us telephone, txt government-diplomatic-representation-in-the-us fax, txt government-diplomatic-representation-in-the-us consulate(s) general, txt government-diplomatic-representation-in-the-us consulate(s), txt government-diplomatic-representation-in-the-us consulate(s), txt government-diplomatic-representation-from-the-us branch office(s), txt government-diplomatic-representation-in-the-us chancery, txt government-diplomatic-representation-from-the-us chief of mission, txt government-diplomatic-representation-from-the-us embassy, txt government-diplomatic-representation-from-the-us mailing address, txt government-diplomatic-representation-from-the-us telephone, txt government-diplomatic-representation-from-the-us fax, txt government-diplomatic-representation-from-the-us consulate(s) general, txt government-diplomatic-representation-from-the-us consulate(s), txt government-constitution, txt government-country-name abbreviation, txt government-government-note, txt government-legal-system, num energy-electricity-access population without electricity, num energy-electricity-installed-generating-capacity, txt military-and-security-military-branches, txt military-and-security-military-service-age-and-obligation, txt military-and-security-maritime-threats, txt military-and-security-military-note, txt communications-communications-note, num transportation-airports, num transportation-airports-with-paved-runways total, num transportation-airports-with-paved-runways 2,438 to 3,047 m, txt transportation-ports-and-terminals major seaport(s), txt transportation-ports-and-terminals oil terminal(s), txt transportation-ports-and-terminals cruise port(s), num transportation-airports-with-paved-runways under 914 m, num transportation-airports-with-unpaved-runways total, num transportation-airports-with-unpaved-runways under 914 m, num transportation-roadways total, num transportation-roadways paved, num transportation-roadways unpaved, num transportation-merchant-marine total, txt transportation-merchant-marine by type, num transportation-airports-with-paved-runways over 3,047 m, num transportation-airports-with-paved-runways 1,524 to 2,437 m, num transportation-airports-with-paved-runways 914 to 1,523 m, num transportation-airports-with-unpaved-runways over 3,047 m, num transportation-airports-with-unpaved-runways 2,438 to 3,047 m, num transportation-airports-with-unpaved-runways 1,524 to 2,437 m, num transportation-airports-with-unpaved-runways 914 to 1,523 m, num transportation-heliports, num transportation-waterways, txt transportation-ports-and-terminals river port(s), num transportation-railways total, num transportation-railways standard gauge, num transportation-railways narrow gauge, num transportation-railways broad gauge, sum transportation-ports-and-terminals oil terminal(s), sum geography-environment-international-agreements signed, but not ratified, sum transnational-issues-refugees-and-internally-displaced-persons refugees (country of origin), sum people-and-society-major-infectious-diseases water contact diseases, sum government-dependent-areas, sum transportation-ports-and-terminals river port(s), sum transnational-issues-refugees-and-internally-displaced-persons idps, sum government-international-organization-participation, sum people-and-society-ethnic-groups, sum people-and-society-major-infectious-diseases animal contact diseases, sum people-and-society-major-infectious-diseases respiratory diseases, sum geography-natural-resources, sum terrorism-terrorist-groups-home-based, sum transportation-ports-and-terminals lng terminal(s) (import), sum geography-environment-international-agreements party to, sum transportation-ports-and-terminals major seaport(s), sum terrorism-terrorist-groups-foreign-based, sum people-and-society-major-infectious-diseases degree of risk, sum transportation-ports-and-terminals lng terminal(s) (export), sum transnational-issues-refugees-and-internally-displaced-persons stateless persons, amount government-international-organization-participation, sum transportation-ports-and-terminals container port(s) (teus), amount transportation-ports-and-terminals container port(s) (teus), sum transportation-pipelines, amount geography-land-boundaries-border-countries-overall, sum transportation-merchant-marine by type, amount military-and-security-military-branches, amount government-dependent-areas |
| <u>Notes</u><br>*There is a high possibility that there are more, undiscovered, MNAR columns in the dataset.<br>**Columns generated from the above inherit the MNAR assumption. |

## Appendix E. Columns Generated from Textual - Assumptions

| Original Column | Assumptions/Operations Performed |
|---|---|
| txt geography-climate | Climates were converted to labels on the basis of the Koppen Climate Classification system (Beck, et al., 2018). Climates not originally included were manually added using external sources. For details regarding assumptions see the original code. |
| txt transportation-pipelines | "condensate/gas" and "oil/condensate" were grouped as "condensate". "refined petroleum product", "oil and refined products", "petroleum products" and "refined petroleum products" were classified as "refined products". natural gas", "gas transmission pipelines", "high-pressure gas distribution pipelines" , "mid- and low-pressure gas distribution pipelines" , "domestic gas" and "gas transmission pipes" were classified as “gas”. "crude oil" and "extra heavy crude" were considered simply as “oil”. "gas and liquid petroleum" was transformed to "liquid petroleum gas". "distribution pipes, "unknown" , "water", "cross-border pipelines" were grouped under the label "oil/gas/water". "ethanol/petrochemical" was labeled as "chemicals". Columns for each type were created and filled with the respective values, otherwise considered MNAR. |
| txt military-and-security-military-service-age-and-obligation | For column "lbl military-and-security-conscription", "no conscription" "no compulsory" were assumed as “no” label while “compulsory” was assumed as “yes”. If an age younger than 15 was found then it was considered that "lbl military-and-security-military-service-age" is “none”. |
| txt military-and-security-military-branches | If "no regular" then "amount military-and-security-military-branches" was assumed 0. |
| txt government-dependency-status | Were generalized into “self-sovereign” and “dependent”. |
| txt government-government-type | "unresolved" was assumed to be "in transition". If “totalitarian” or “dictatorship” was found in the CIA comments of the subfield then it was adopted as government type. |
| txt government-legal-system | "local" was grouped with "customary". "islamic" was grouped with "religious". |
| txt government-executive-branch head of government | "head of government" was labeled as “prime minister”. |